\documentclass[preprint]{ptephy_v1}

\preprintnumber{XXXX-XXXX} 
\usepackage{hyperref}

\usepackage{graphicx} 
\usepackage{color}
\usepackage{siunitx}

\usepackage{mathtools}
\usepackage{enumerate}
\usepackage{lineno}
\usepackage[normalem]{ulem}
\usepackage{multicol}
\usepackage{url}
\usepackage{subcaption}
\usepackage{cleveref}

\crefname{figure}{Fig.}{Figs.}
\crefname{equation}{Eq.}{Eqs.}

\begin{document}

\title{Broadband, Dead-Time-Free Spectrometer Using RFSoC for WISP Dark Matter Searches}
\author{Hiroki Takeuchi}
\affil{Department of Physics, Faculty of Science, Kyoto University, Kyoto 606-8502, Japan \email{takeuchi.hiroki.74w@st.kyoto-u.ac.jp}}
\author[1]{Junya Suzuki}
\author[2, 1]{Shunsuke Adachi}
\affil{Hakubi Center for Advanced Research, Kyoto University, Kyoto 606-8501, Japan}
\author[1]{Osamu Tajima}


\begin{abstract}%

Ultra-light dark matter candidates, such as axions and dark photons, particularly within the mass range of $\mu\si{eV}$ to $\si{meV}$, have garnered significant research interest.
However, the effectiveness of existing experiments is often hindered by the narrow bandwidths of conventional spectrometers.
This study presents a novel spectrometer solution based on a RFSoC, which integrates a high-speed data converter, field programmable gate array (FPGA), and a CPU within a single chip.
The spectrometer’s programmable logic implements an optimized fast Fourier transform (FFT) that minimizes memory usage, enabling an instantaneous bandwidth of \SI{4}{GHz} with a frequency resolution of \SI{16}{kHz}. 
This configuration achieves more than 99.6\% time efficiency, effectively doubling search efficiency, compared with traditional approaches.
This paper provides a comprehensive description of the hardware architecture, FPGA firmware, and software implementation, with a focus on the resource-efficient pipelined FFT algorithm.
Performance validation results are also presented, demonstrating the accuracy and time efficiency of the developed spectrometer, highlighting its potential for advancing dark matter search.

\end{abstract}

\subjectindex{xxxx, xxx}

\maketitle

\section{Introduction}
Dark matter constitutes $26.8\%$ of the total energy density of the universe~\cite{2020}, yet its nature remains elusive.
Understanding dark matter is a crucial issue in both cosmology and particle physics.
Owing to the limited knowledge about the properties of dark matter, researchers are exploring a wide range of potential candidates with varying mass scales for the particle.
In recent years, a group of particles, known as weakly interacting slim particles has attracted attention as a dark matter candidate~\cite{PaolaArias2012}, 
including axions and dark photons.

Axions and dark photons can be converted into ordinary photons, allowing for experiments aimed at detecting dark matter candidates.
The dish-antenna method proposed in Ref. \cite{Horns2013} enables the exploration of a broad range of particle masses.
The experimental setup involves a mirror (with a magnet for axion searches), receiver circuit, and spectrometer.
The frequency distribution of the conversion photon displays a sharp peak at $\nu_0 = m_\mathrm{DM}c^2 / h$ with a full width at half maximum of $\Delta \nu / \nu_0 \sim 10^{-6}$ when assuming the isothermal halo model for the velocity distribution of dark matter.
Here, $m_\mathrm{DM}$ represents the mass of a particle, and $c$ represents the speed of light. $h$ denotes the Plank’s constant.
Therefore, the detection of axions or dark photons can be achieved by observing the signal spectrum received by a spectrometer and identifying this spectral peak.
The dark photon and axion are characterized by two main parameters: their mass and coupling constant with photons, denoted as $\chi$ and $g_{a\gamma\gamma}$ for the dark photon and axion, respectively.
When a spectral peak is detected, the mass $m_{DM}$ can be determined by the peak frequency, whereas the coupling constant $\chi$ or $g_{a\gamma\gamma}$ can be determined by the intensity of the conversion photon.
Past studies~\cite{suzuki2015, Tomita2020, Kotaka2023, Adachi2024, Bajjali2023, Knirck2024} utilizing the dish-antenna method have achieved direct searches with a sensitivity that surpasses the indirect constraints of cosmology~\cite{PaolaArias2012, Witte2020}.

Among these studies, Refs. \cite{Adachi2024, Kotaka2023} did not fully utilized the wide-band nature of the dish-antenna method, owing to the limited bandwidth of the spectrometers employed.
They instead employed commercial signal analyzers, which restricted their study to a narrow bandwidth of \SI{2}{MHz}.
Conversely, an antenna and radio-frequency (RF) circuit can cover a wide frequency range of $O(\si{GHz})$, allowing for simultaneous searches across the full spectrum. For these experiments, a spectrometer capable of analyzing over \SI{1}{GHz} bandwidth simultaneously is needed.
Notable attempts to develop such a wide-band spectrometer include Fourier transform versions that require graphical processing units (GPUs) or field-programmable gate arrays (FPGAs).
For example, in Ref. \cite{Bajjali2023}, a GPU was used to analyze a \SI{4}{GHz} band with a time efficiency of $60\%$.
Notably, commercial FPGA-based spectrometers can provide dead-time-free analyses over a \SI{2.5}{GHz} bandwidth~\cite{Klein2012}.

This study presents a dead-time free spectrometer with \SI{4}{GHz} bandwidth. The spectrometer is built on an RFSoC 2x2 board, utilizing an RFSoC chip that integrates high-speed data converters, FPGA, and CPU~\cite{RFSoC}. The performance evaluation of this spectrometer is also discussed in detail.

\section{Design Strategy and Implementation Method}
\label{sec-strategy}

\subsection{Requirements}
\label{sec_req}
One key specification that directly impacts search efficiency is called the ``instantaneous bandwidth,'' which refers to the frequency bandwidth around the center frequency that can be acquired in a single measurement.
The upper limit of the instantaneous bandwidth is determined by the sampling rate of the analog-to-digital converter (ADC), responsible for converting analog signals to digital signals.
Given the bandwidth capabilities of current high-speed ADCs, our goal was to achieve a bandwidth in the range of several \si{GHz}.

Moreover, efficient signal processing is essential to maximize time efficiency and must be performed with no dead time.

Additionally, achieving a high-frequency resolution is crucial for identifying spectral peaks.
To search for signals around \SI{30}{GHz}, a frequency resolution of \SI{30}{kHz} is required. 
This specific frequency range has not been explored in previous studies on axion or dark photon searches.
With an instantaneous bandwidth of \SI{4}{GHz}, dividing the band into $1.3 \times 10^5$ points is necessary to achieve the desired resolution.

\subsection{Hardware Design}
To satisfy the above requirements, a high-speed ADC capable of several giga-samples per second (GSPS) is needed, along with a high-performance FPGA capable of processing the ADC output.
For this, we employed an AMD Xilinx chip known as the RFSoC~\cite{RFSoC}, which integrates a CPU, FPGA, ADCs, DACs, and other RF circuits into a single system on chip (SoC).
Additionally, the processing system (PS) housing the CPU and the programmable logic (PL) housing the FPGA were both connected to external dynamic random access memory (DRAM) chips.

\begin{figure}
   \centering
   \includegraphics[width=4.1in]{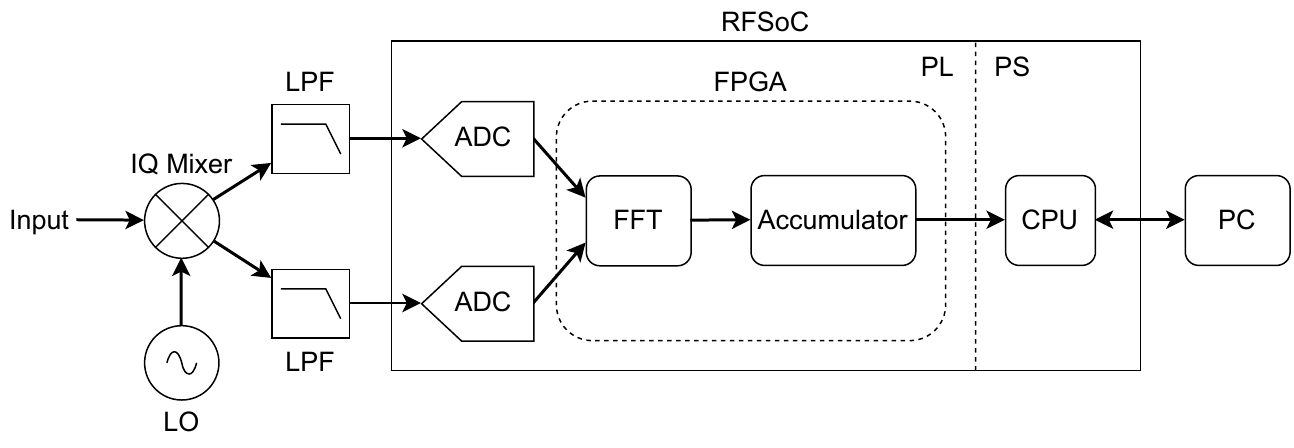}
   \includegraphics[width=1.8in]{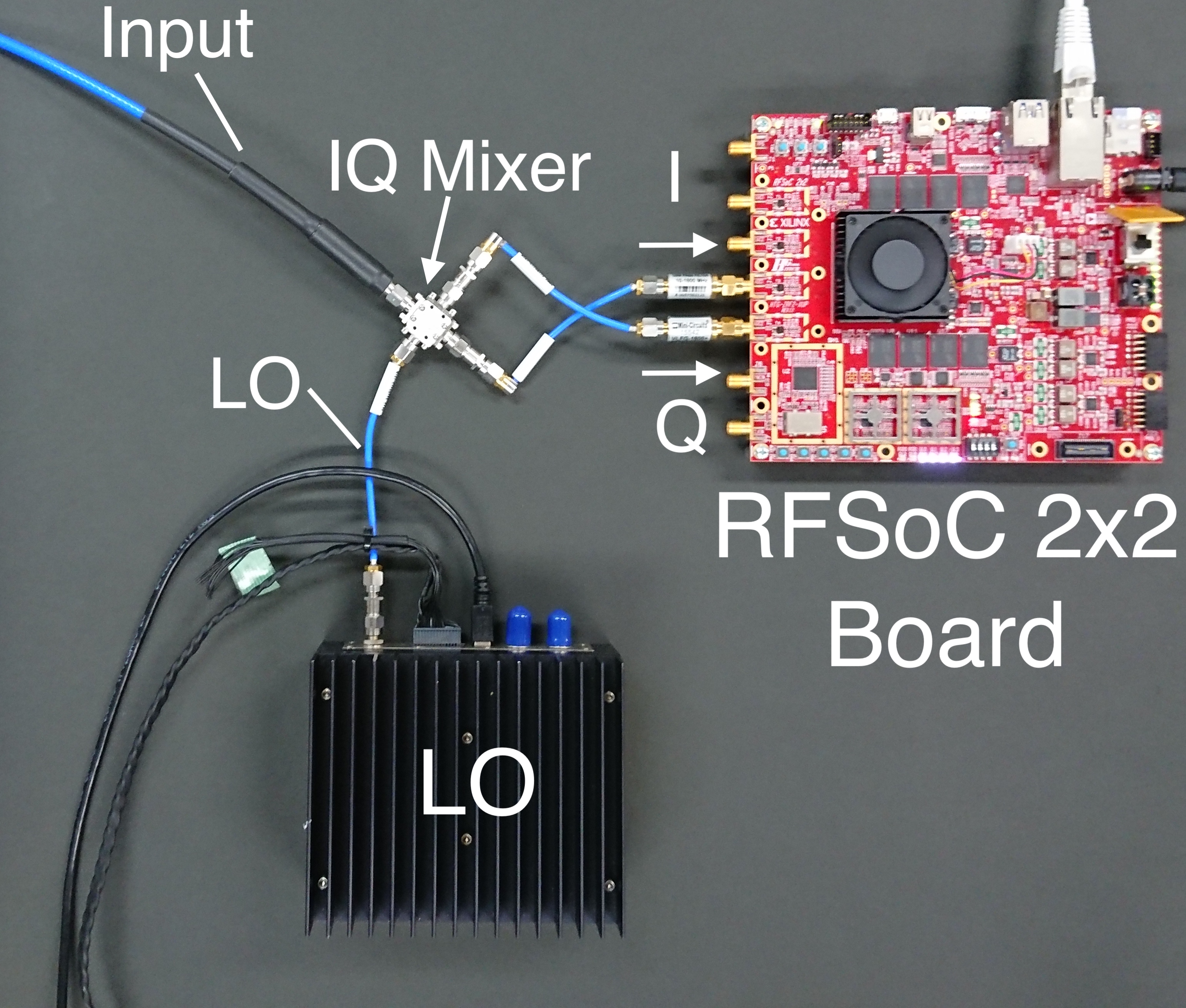}
   \caption{(Left) Schematic of RF circuits and digital signal processing. The input signal is down-converted by utilizing the IQ mixer and sent to the RFSoC. There, the down-converted signals are converted to digital time-series signals by the \SI{4}{GSPS} ADCs, and the power spectrum is obtained using the parallel FFT circuit on the FPGA. An accumulator circuit calculates the time average, with the obtained power spectrum transferred to the embedded CPU, which is then sent to an external computer over Ethernet. (Right) Photo of our proposed spectrometer.}
   \label{fig_setup_schematic}
\end{figure}

An overview of the spectrometer design and a photograph of the developed spectrometer setup is shown in \Cref{fig_setup_schematic}.
Detailed descriptions of each block will be provided in the following sections.
The spectrometer components were selected to target a frequency of approximately \SI{30}{GHz}, and a Marki microwave MMIQ-1040LS was utilized as the IQ mixer.
Notably, with different IQ mixer models, the spectrometer can be utilized for a variety of frequency bands.
A HiTech Global RFSoC 2x2 board~\cite{RFSoC2x2} equipped with the ZYNQ UltraScale+ RFSoC XCZU28DR, was used, featuring two 12-bit \SI{4.096}{GSPS} ADCs,
and a \SI{1}{Gbps} Ethernet cable was used for external communications.
ADC synchronization was achieved using the reference clock on the RFSoC board.
The IQ mixer and ADCs are connected through SMA cables, with a low-pass filter (Mini-Circuits VLFG-1800+) inserted between them to limit the signal to below \SI{2}{GHz}.

To detect spectral peaks with a width of approximately $10^{-6}$ times \SI{30}{GHz}, we set the FFT size to $N = 2^{17}$ to achieve a frequency resolution of \SI{31.25}{kHz}\footnote{$\SI{4.096}{GHz}$ bandwidth $/ 2^{17} = \SI{31.25}{kHz}$}.
Subsequently, we developed a high-resolution version with an FFT size of $N = 2^{18}$, providing a frequency resolution of \SI{15.625}{kHz}.

\begin{figure}
   \centering\includegraphics[width=3.5in]{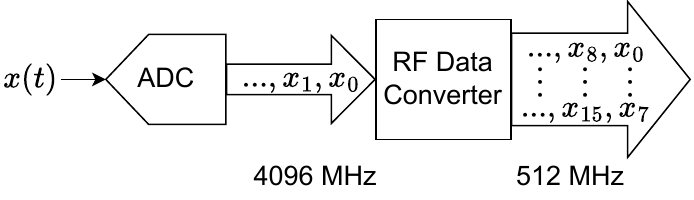}
   \caption{Signal handling by the RF data converter. The ADC receives data at \SI{4.096}{GSPS} and transmits it to the RF data converter on the FPGA. The RF data converter operates at \SI{512}{MHz}, which corresponds to $1/8$ of the ADC clock frequency, and outputs data from eight time points per clock cycle.}
   \label{fig_rfdc}
\end{figure}

The data acquired by the ADCs were streamed to the PL through the RF data converter’s intellectual property (IP) core, whose output was \SI{512}{MHz} with eight time points per clock cycle (\Cref{fig_rfdc}).

\subsection{Firmware Design}
\subsubsection{Clock Frequency Conversion and FFT}
\label{sec_des_fft}
Operating signal processing at \SI{512}{MHz} presents challenges in satisfying PL-resource timing constraints.
To address this issue, we utilized input distribution and asynchronous first-in-first-out (FIFO) logic to increase the number of data lanes to 16 while reducing the clock frequency by half.

\begin{figure}
   \centering
   \includegraphics[width=5in]{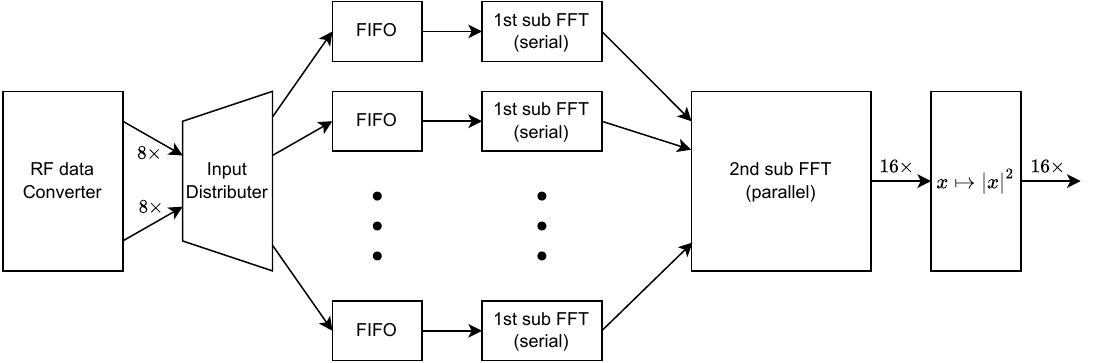}
   \caption{Block diagram of the FFT block. Signal processing is performed by four types of IPs from left to right: input distribution circuit (asynchronous FIFO), first sub-FFT circuit (serial FFT), second sub-FFT (parallel FFT), and power spectrum calculation circuit.}
   \label{fig_arch_16}
\end{figure}

Subsequently, ADC signals were converted into a power spectrum using an FFT block.
Further details regarding the FFT circuit can be found in \Cref{sec-fft}.
Following the FFT computation, the power spectrum was determined from the resulting complex spectrum.
The block diagram is shown in \Cref{fig_arch_16}.
The input distribution circuit, secondary sub-FFT circuits, and power spectrum calculation circuits were implemented through high-level synthesis (HLS), which involves converting C++ code into a hardware description language.
A Vitis HLS was used, and existing IPs were integrated for clock conversions and initial sub-FFT computations.

To align the output to the accumulator circuit, it was configured as a \SI{54}{bit} fixed-point number for the $N = 2^{17}$ FFT circuit specification or a \SI{32}{bit} floating-point number for the $N = 2^{18}$ FFT circuit specification.
These specifications are listed in \Cref{tab_fft_spec}.

\begin{table}
\centering
\caption{Specifications of the FFT block.}
\begin{tabular}{lrr}
\hline
\hline
Specification & $N = 2^{17}$ & $N = 2^{18}$ \\
\hline
Parallel channels & 16 & 16 \\
Input format & Fixed-point & Fixed-point \\
Input bit width & 12 bits & 12 bits \\
Format at FFT completion & Fixed-point & Fixed-point \\
Bit width at FFT completion & 29 bits & 30 bits \\
Output format & Fixed-point & Floating-point \\
Output bit width & 54 bits & 32 bits \\
\hline
\hline
\end{tabular}
\label{tab_fft_spec}
\end{table}

\subsubsection{Accumulator Block}
\label{sec_des_acc}

\begin{figure}
    \centering
    \includegraphics[width=5in]{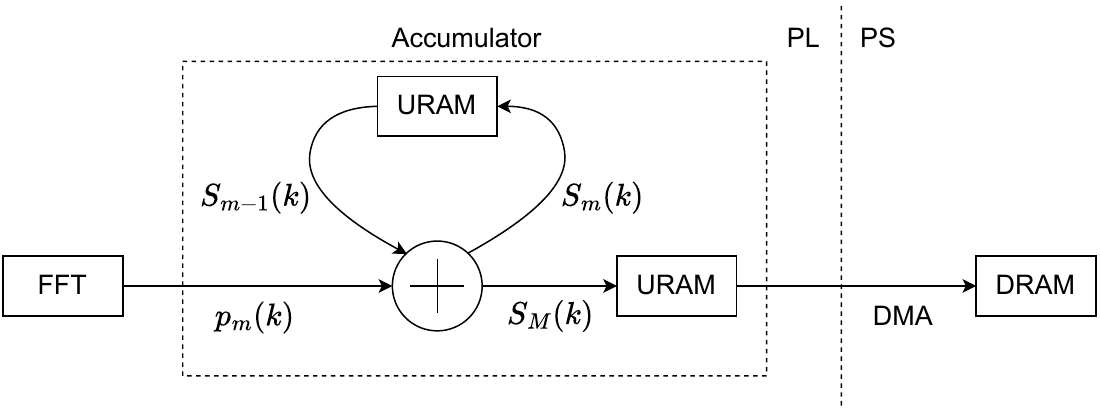}
    \caption{Overview of the accumulator circuit. The time average of the power spectrum output from the FFT circuit was calculated on the FPGA and transferred to the DRAM in the PS through DMA for processing by the CPU.}
    \label{fig_acc}
\end{figure}

An accumulator circuit calculates the time average (sum) of the power spectrum, effectively reducing the data rate before transferring it to the PS.
An overview of the accumulator circuit is shown in \Cref{fig_acc}.
Initially, the power spectrum was fed from the FFT circuit, with the partial sum temporarily stored in the UltraRAM (URAM) on the PL.
After accumulating data from $M$ iterations, the final result was then transferred to a data buffer, constructed using the URAM resource.
This entire process was performed simultaneously for $16$ values at a frequency index ok $k$.

For a set averaging count of $M=1024$ and an $N=2^{17}$ FFT circuit, the input width is \SI{54}{bit}, and the output width is \SI{64}{bit}.
However, for an FFT circuit with $N=2^{18}$, both input and output utilize \SI{32}{bit} floating-point numbers (\Cref{tab_acc}).
Notably, the accumulator circuit was implemented entirely through high-level synthesis.

\begin{table}
\centering
\caption{Specifications of the accumulator circuit.}
\begin{tabular}{lrr}
\hline
\hline
Specification & $N = 2^{17}$ & $N = 2^{18}$ \\
\hline
Averaging count $M$      & $1024$ & $1024$ \\
Input format & Fixed-point & Floating-point \\
Input bit width & \SI{54}{bit} & \SI{32}{bit} \\
Output format & Fixed-point & Floating-point \\
Output bit width & \SI{64}{bit} & \SI{32}{bit} \\
\hline
\hline
\end{tabular}
\label{tab_acc}
\end{table}

\subsubsection{Resource Usage}
\label{sec_des_imp}
The total amount of resources in the FPGA of the RFSoC and the amount of resources utilized are listed in \Cref{tab_util}.
Apart from the URAM, which stores FFT results in the accumulation block, each resource type was utilized at less than 50\%, which allows for expansion through parallelization, as well as increasing the FFT size.
URAM was primarily allocated for storing FFT results in the accumulation block,
and its resource consumption could be minimized by decreasing the arithmetic precision of the stored data.
We successfully implemented firmware with FFT circuits with $N = 2^{17}$ at \SI{64}{bit} fixed-point numbers and $N = 2^{18}$, with precision limited to \SI{32}{bit} floating-point numbers.
Our spectrometer specifications are listed in \Cref{tab_fft}.

\begin{table}[]
\centering
\caption{FPGA resource utilization of the spectrometer}
\begin{tabular}{lrrrrr}
\hline
\hline
Resource Type & \multicolumn{2}{c}{Utilization}            & \multicolumn{2}{c}{Usage}                                       & Available            \\
              & \multicolumn{1}{l}{$N=2^{17}$} & $N=2^{18}$ & \multicolumn{1}{l}{$N=2^{17}$} & \multicolumn{1}{l}{$N=2^{18}$} & \multicolumn{1}{l}{} \\ \hline
LUT           & $20.0\%$                      & $25.5\%$  & $\SI{85}{k}$                          & $\SI{108}{k}$                         & $\SI{425}{k}$               \\
FF            & $18.5\%$                      & $20.7\%$  & $\SI{157}{k}$                         & $\SI{176}{k}$                         & $\SI{851}{k}$               \\
BRAM          & $14.5\%$                      & $24.8\%$  & $157$                        & $268$                          & $1080$               \\
URAM          & $77.5\%$                      & $80.0\%$  & $62$                           & $64$                           & $80$                 \\
DSP           & $15.8\%$                      & $15.9\%$  & $674$                          & $678$                          & $4272$               \\ \hline
\hline
\end{tabular}
\label{tab_util}
\end{table}

The FFT circuit achieved a throughput surpassing the ADC sampling rate, enabling the full \SI{4.096}{GHz} bandwidth of the ADC to be utilized.
Additionally, a resolution of \SI{15.6}{kHz} was achieved, which is suitable for the target signal at approximately \SI{30}{GHz}.

\begin{table}
\centering
\caption{Summary of specifications of the developed spectrometer}
\begin{tabular}{ll}
\hline
\hline
Item       & Performance Value \\
\hline
Frequency Bandwidth       & $\SI{4.096}{GHz}$ \\
Input & Complex (IQ) \\
FFT Size        & $2^{17}$ / $2^{18}$ \\
Frequency Resolution      & $\SI{31.25}{kHz}$ / $\SI{15.625}{kHz}$ \\
ADC Resolution & 12 bit \\
ADC Sampling Rate & $\SI{4.096}{GSPS}$ \\
FFT Circuit Throughput & $\SI{4.800}{GSPS}$ \\
Time Averaging Interval & $\SI{32.768}{ms}$ / $\SI{65.536}{ms}$ \\
\hline
\hline
\end{tabular}
\label{tab_fft}
\end{table}

\subsection{Software Design}
\label{sec-software}

\begin{figure}
   \centering\includegraphics[width=5in]{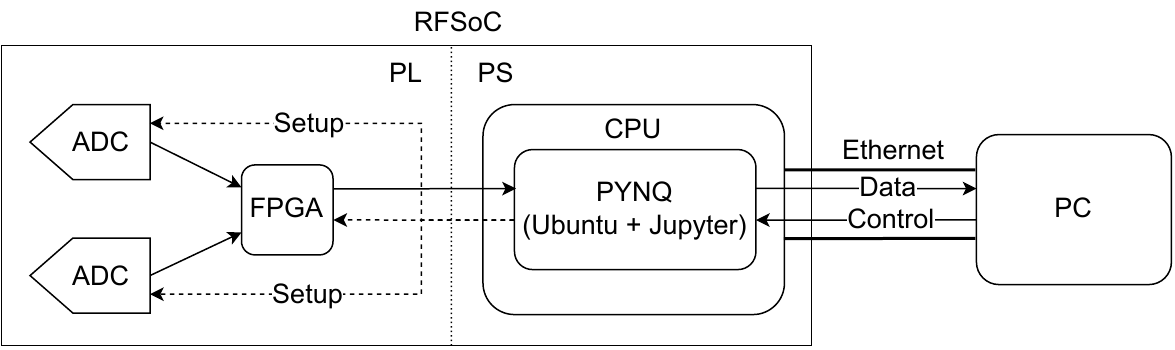}
   \caption{Overview of the software component. Python scripts were run on PYNQ to perform the initial setup of the FPGA and ADCs, clock synchronization settings for the two ADCs, and transfer of the spectrum obtained by the FPGA to the PC.}
   \label{fig_soft}
\end{figure}

An overview of the interface between the firmware and software is shown in \Cref{fig_soft}.
Data were reordered from bit-reverse to natural order after accumulation on the PL and subsequently transferred to the PS's DRAM through an Advanced eXtensible Interface (AXI) port utilizing Xilinx’s direct memory access (DMA) core.
To facilitate physical routing during implementation, the DMA circuit operated at \SI{100}{MHz}.

Subsequently, the data were time-averaged in the PS and transmitted to an external PC through a gigabit Ethernet cable on the PS side using the ZeroMQ library~\cite{ZeroMQ}.
The PYNQ framework~\cite{PYNQ} was employed for PL reconfiguration, data transmission, and post-processing.

On the external PC, a standard Python script was utilized to communicate with the RFSoC and to store the received data.

\section{Resource-Efficient Implementation of the FFT Circuit on FPGA}
\label{sec-fft}
This section describes our resource-efficient implementation methods for pipelined FFT circuits.
These methods were utilized in the design of our spectrometer to minimize circuit area, ensuring compatibility with the limited resources available in the PL of the RFSoC chip.
In the implementation of the FFT circuit, the rotation circuit must be efficient, as it demands significant circuit resources and consumes a substantial amount of memory (e.g., for complex multiplication with several coefficients).
To reduce memory consumption, we divided the rotation process into two stages and employed a Radix-$2^4$ architecture~\cite{OH_2005, 6118316}.

\subsection{Basics of the Pipelined FFT and the Terminology}
\label{sec_fft_method}

In an FFT-based spectrometer, the time-series input data, $x(t)$, are converted into a frequency spectrum, $\tilde{x}(k)$, through a discrete Fourier transform (DFT):
\begin{align}
    \tilde{x}(k) = \sum_{t = 0}^{N - 1} W_N^{kt} x(t),
    \label{eq_DFT}
\end{align}
where $N$ represents the length of the DFT, and $k, t$ assume integer values from $0$ to $N - 1$.
$W_N$ is defined as $W_N = e^{-2\pi i \frac{1}{N}}$, representing the $N$th primitive root of unity.
The Cooley-Tukey algorithm~\cite{Cooley1965}, generally referred to as the FFT, reduces the computational cost of DFT from $O(N^2)$ to $O(N\log N)$ by recursively breaking down the DFT into smaller components. For $N = 2^n$, $\tilde{x}(k)$ is described as follows:
\begin{align}
    \notag
    \tilde{x}\left(k\right) = &\sum_{t_0 = 0}^1 W_2^{k_{n - 1}t_0}
    W_{2^n}^{k_{0:n-2} t_0}
    \sum_{t_1 = 0}^1 W_2^{k_{n - 2}t_1}
    W_{2^{n - 1}}^{k_{0:n-3} t_1} \\
    &\cdots
    \sum_{t_{n - 2} = 0}^1 W_2^{k_1t_{n-2}}
    W_{2^2}^{k_{0} t_{n - 2}}
    \sum_{t_{n-1} = 0}^{1} W_2^{k_0t_{n-1}} x(t).
    \label{eq_FFT_nd}
\end{align}
Here, for a non-negative integer $x$ represented in binary notation, $x_i$ denotes its $i$th digit from the right, whereas $x_{i:j}$ represents the number from the $i$th to $j$th digits. For example, if $x = 6$ (binary: $110^{(2)}$), then $x_0 = 0, x_{1:2} = 3$.

The concept of pipelined FFT, which implements FFT as a continuous pipelined flow in a circuit, was proposed~\cite{OLeary1970, Groginsky1970} and has been widely utilized in achieving high throughput.
Various circuit architectures have been proposed to cater to a wide range of applications~\cite{Garrido2021}.

\begin{figure}
   \centering
   \includegraphics[width=5.5in]{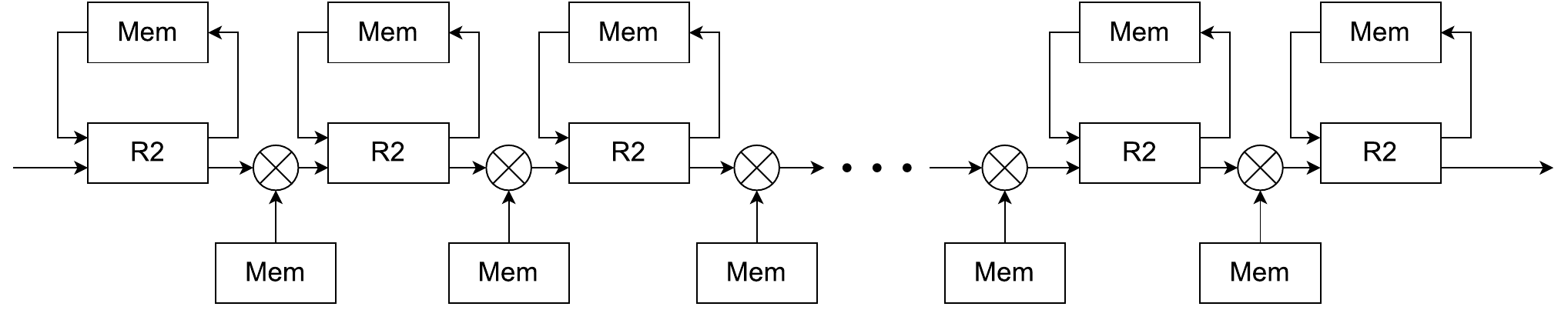}
   \caption{Structure of the SDF pipeline circuit. ``R2'' denotes a Radix-2 butterfly operation, and ``Mem'' represents a memory for temporary data storage or a table for rotation operation.}
   \label{fig_FFT_SDF_R2}
\end{figure}
\begin{figure}
   \centering
   \includegraphics[width=5.5in]{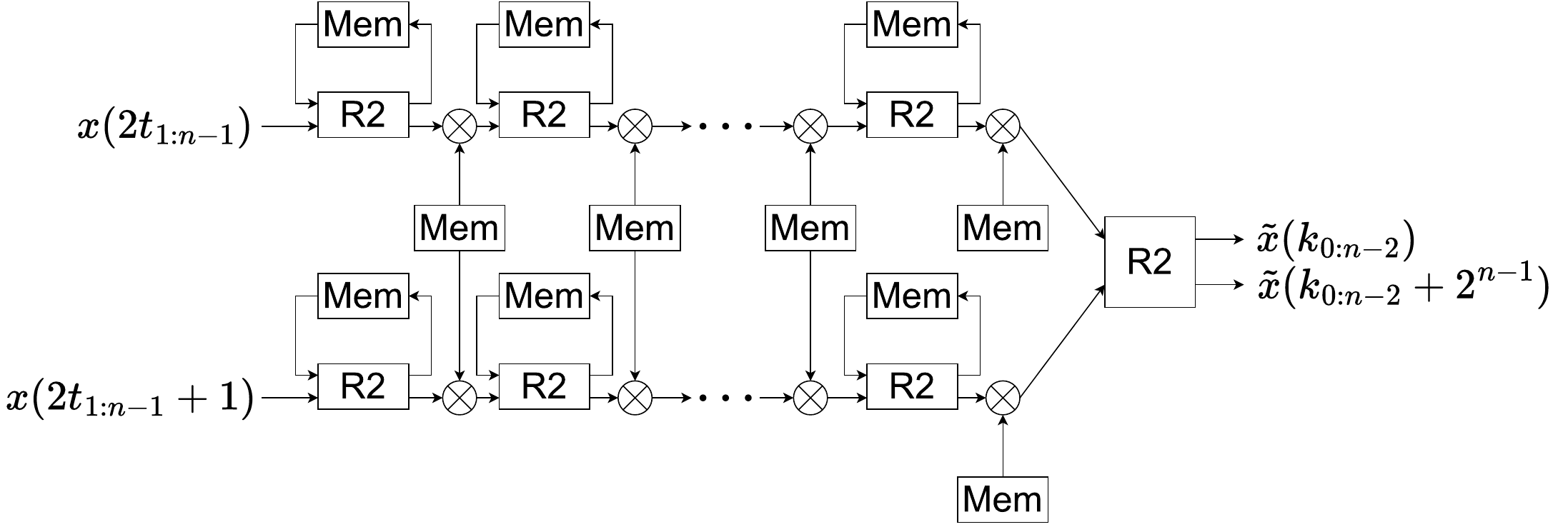}
   \caption{Two-parallel MDF architecture comprising two SDF circuits and a Radix-2 butterfly unit, achieving doubled throughput.}
   \label{fig_FFT_SDF_P2}
\end{figure}
The schematic of the single-path delay feedback (SDF) architecture of the pipelined FFT is shown in Fig.~\ref{fig_FFT_SDF_R2},
whereas the multi-path delay feedback (MDF) architecture, which can achieve twice or even higher throughput, is shown in Fig.~\ref{fig_FFT_SDF_P2}.
By examining the twiddle factors $W_2(=-1)$ and $W_{2^{k}}$ in \Cref{eq_FFT_nd},
the FFT can be expressed as a combination of 2-point FFT, which transforms input $A$ and $B$ into $A+B$ and $A-B$, along with a general rotation by the power of $W_{2^{k}}$.
The former process is known as a (Radix-2) butterfly operation (``R2'' in the figures).
The latter is typically implemented using a look-up table (LUT), where pre-computed twiddle factors are stored and retrieved for multiplication at runtime.
Although this approach requires $O(2^k)$ memory size for computing powers of $W_{2^k}$, it can be implemented using a single complex multiplier.
In the SDF architecture, which processes one input per clock cycle, temporary memory is utilized to store input data so that the butterfly operation can be executed using subsequent inputs.
The 2-parallel MDF architecture shown in Fig.~\ref{fig_FFT_SDF_P2} features two SDF circuits in parallel, each handling even- and odd-indexed data input.
These circuits merge their outputs at the end to produce the upper and lower halves of the frequency domain data simultaneously.

\subsection{Resource Reduction Techniques for Pipelined FFT}

\subsubsection{Memory Reduction of the Rotation Unit by Splitting the Twiddle Factor Table}

The memory requirements for the rotation process, shown in \Cref{eq_FFT_nd} and \Cref{fig_FFT_SDF_R2}, are directly proportional to the FFT length.
For an FFT length of $N = 2^n \quad (n = 17 \text{ or } 18)$, approximately \SI{10}{Mbit} of memory is required to implement a rotation.
To facilitate simultaneous access to the twiddle factor table at each stage, distributed memory resources, such as block RAM (BRAM) on FPGA are used to enable high-bandwidth random access.
Given the limited availability of BRAM resources, we optimized memory utilization for the rotation unit by partitioning the twiddle factor table.

First, the size of the twiddle factor table was reduced by factoring $W_{4}$ from general $W_N$:
\begin{align*}
   W_N^m = W_4^{m_{n-2:n-1}} W_N^{m_{0:n-3}},
\end{align*}
because multiplication of $W_{4} = i$ or its power can be easily performed using a sign flip and swapping real and imaginary components.

\begin{figure}
   \centering
   \includegraphics[width=3in]{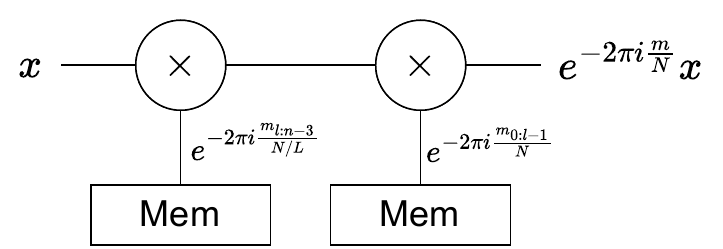}
   \caption{Overview of the developed split LUT rotation unit. By splitting the rotation into two components, the number of complex multiplications doubled compared with those of the conventional LUT rotation unit; however, the rotation memory utilization was significantly reduced from $O(N)$ to $O(\sqrt{N})$.}
   \label{fig_rot_dev}
\end{figure}

To further minimize memory utilization, we set $L = 2^l (< N)$ and decomposed it into a product of three twiddle factors:
\begin{align}
   W_N^m = W_4^{m_{n-2:n-1}} W_\frac{N}{L}^{m_{l:n-3}} W_N^{m_{0:l-1}},
   \label{eq_rot_dev}
\end{align}
where $W_\frac{N}{L}^{m_{l:n-3}}$ represents a coarse rotation and $W_N^{m_{0:l-1}}$ represents a fine rotation.
Because $m_{0:l-1}$ is an integer in the range $[0, L)$,
$2L$ real numbers are required to store $W_N^{m_{0:l-1}}$.
Similarly, because $m_{l:n-3}$ is an integer in the range [0, $\frac{N}{4L})$, $2 \times \frac{N}{4L}$ real numbers are sufficient for storing $W_\frac{N}{L}^{m_{l:n-3}}$.
Moreover, by leveraging the symmetry inherent in cosine and sine functions, the memory requirement for $W_\frac{N}{L}^{m_{l:n-3}}$ can be reduced to $\frac{N}{4L}$ real numbers.
Hence, through the computation of \Cref{eq_rot_dev} using the circuit depicted in \Cref{fig_rot_dev}, the rotation operation can be executed utilizing a total memory of only $2L + \frac{N}{4L}$ real numbers.
This rotation unit configuration is henceforth referred to as a ``split LUT rotation unit.''

The memory usage of the split LUT rotation unit is minimized when, for $N = 2^n$, $L = 2^\frac{n - 3}{2}$ if $n$ is odd, or $L = 2^\frac{n - 4}{2}, 2^\frac{n - 2}{2}$ if $n$ is even.
This memory requires $\sqrt{2N}$ real numbers for odd $n$, and $\frac{3}{2} \sqrt{N}$ real numbers for even $n$.
Because the required memory for a conventional rotation unit is $N / 4$ real numbers, our memory usage is reduced by an order of magnitude.
For example, $N = 2^{17}$ can be reduced to $1 / 64$.
Alternatively, the number of required arithmetic units (e.g., multipliers) is approximately doubled because the rotation operation must be performed twice.
To address this tradeoff, we used the split LUT rotation only for the rotation units requiring the largest LUT size.
These units require sixteen times more LUT size than the second largest ones.

\subsubsection{Radix-$2^4$ MDF Architecture}
As demonstrated, LUT splitting reduces the memory consumption in the general rotation units.
Furthermore, the number of general rotation units can be reduced by optimizing the architecture.
Hence, we adopted the Radix-$2^4$ MDF architecture~\cite{OH_2005, 6118316}, characterized by the fewest general rotation units for a 16-parallel FFT circuit.

We split $k$ and $t$ as $k = k_{0:n-5} + 2^{n-4} k_{n-4:n-1},~ t = t_{0:3} + 2^4 t_{4:n-1}$ and computed the results as follows:
\begin{align}
    \tilde{x}\left(k\right) &= \sum_{t_{0:3} = 0}^{15}\sum_{t_{4:n-1} = 0}^{2^{n-4} - 1} W_N^{\left(k_{0:n-5} + 2^{n-4}k_{n-4:n-1}\right)(t_{0:3} + 2^4t_{4:n-1})} x(t) \\
    &= \sum_{t_{0:3} = 0}^{15} W_{2^4}^{k_{n-4:n-1}t_{0:3}} W_N^{k_{0:n-5}t_{0:3}} \sum_{t_{4:n-1} = 0}^{2^{n-4} - 1} W_{2^{n-4}}^{k_{0:n-5}t_{4:n-1}} x(t).
    \label{FFT_16}
\end{align}
The inner sum comprises isolated DFTs of size $\frac{N}{16}$ and the outer sum involves a combination of rotation operations and a size-16 DFT, processed using 15 rotation units and a fully parallelized 16-parallel FFT circuit.
Within this circuit, we split $k_{n-4:n-1}$ and $t_{0:3}$, and computed the results as follows:
\begin{align}
    \notag
    \sum_{t_{0:3} = 0}^{15} W_{2^4}^{k_{n-4:n-1}t_{0:3}} &=
        \sum_{t_0 = 0}^{1} W_2^{k_{n-1}t_0}  W_{4}^{k_{n-2}t_0} 
        \sum_{t_1 = 0}^{1} W_2^{k_{n-2}t_1}  W_{16}^{k_{n-4:n-3}t_{0:1}} \\
        &\qquad \sum_{t_2 = 0}^{1} W_2^{k_{n-3}t_2}  W_{4}^{k_{n-4}t_2}
        \sum_{t_3 = 0}^{1} W_2^{k_{n-4}t_3}.
    \label{FFT_2-4}
\end{align}
Among the rotation operations performed in \Cref{FFT_2-4}, the components represented by $W_2^m$ and $W_4^m$ can be expressed through sign inversion, including swapping the real and imaginary parts, which eliminates the need for multipliers.
Moreover, as the values of $k_{n-4:n-3}$ and $t_{0:1}$ are predetermined, the rotation operation $W_{16}^{k_{n-4:n-3}t_{0:1}}$ can be executed using a constant multiplier.
Although $W_{16}$ cannot reduce the number of multiplications, unlike $W_2$ or $W_4$, the cost of constant multiplication is minimal, and no rotation memory is required as the rotation angle is known.

\begin{figure}
    \centering
    \includegraphics[width=6.0in]{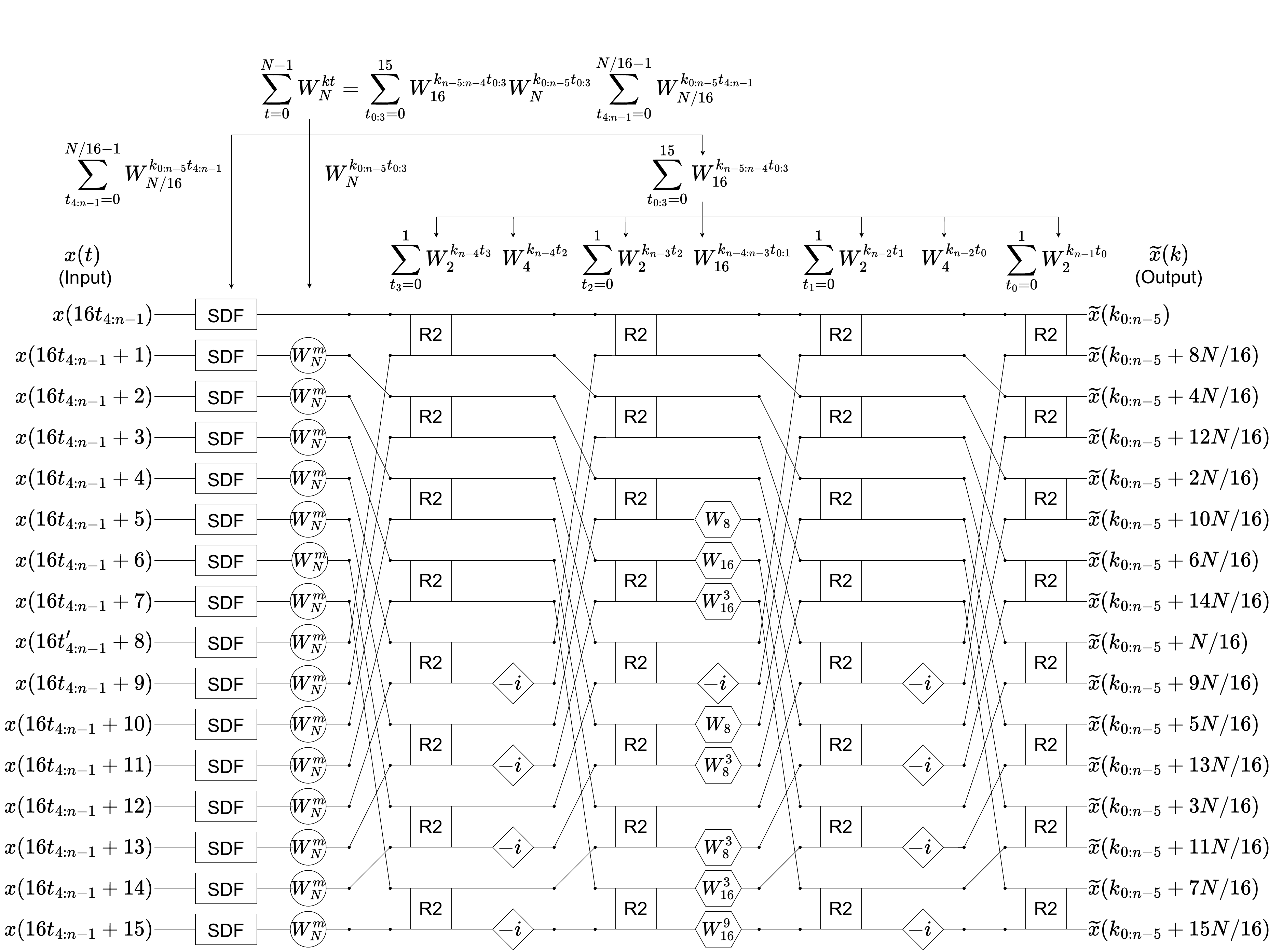}
    \caption{FFT circuit configuration. The calculation was divided into 16 SDF circuits and a 16-parallel MDF FFT circuit. The latter was further divided into a Radix-$2^4$ circuit based on \Cref{FFT_2-4}, comprising Radix-2 butterfly units (R2), $-i$ multipliers (diamond), constant rotation units (hexagon), and general rotation units (circle). The general rotation units employ the split LUT method to reduce memory utilization.}
    \label{fig_arch_R24}
\end{figure}
The configuration shown in \Cref{fig_arch_R24} can be obtained by constructing the circuit from \Cref{FFT_16,FFT_2-4}.
We adopted Xilinx's FFT core for the 16 SDF circuits and a split LUT rotation unit for general rotation following each circuit.
The Radix-$2^4$ architecture reduced the number of general rotation units to 15.
For comparison, the original Radix-2 architecture utilized 32 units, whereas the Radix-$2^2$ architecture, which is widely utilized, requires 24 units.

\section{Performance Evaluation}
We evaluated the performance and linearity of our spectroscopy method using a monotone RF source as input to the spectrometer.
Additionally, we assessed the time efficiency of the spectrometer using a blackbody radiation source.

\begin{figure}
	\centering
  \includegraphics[width=3.5in]{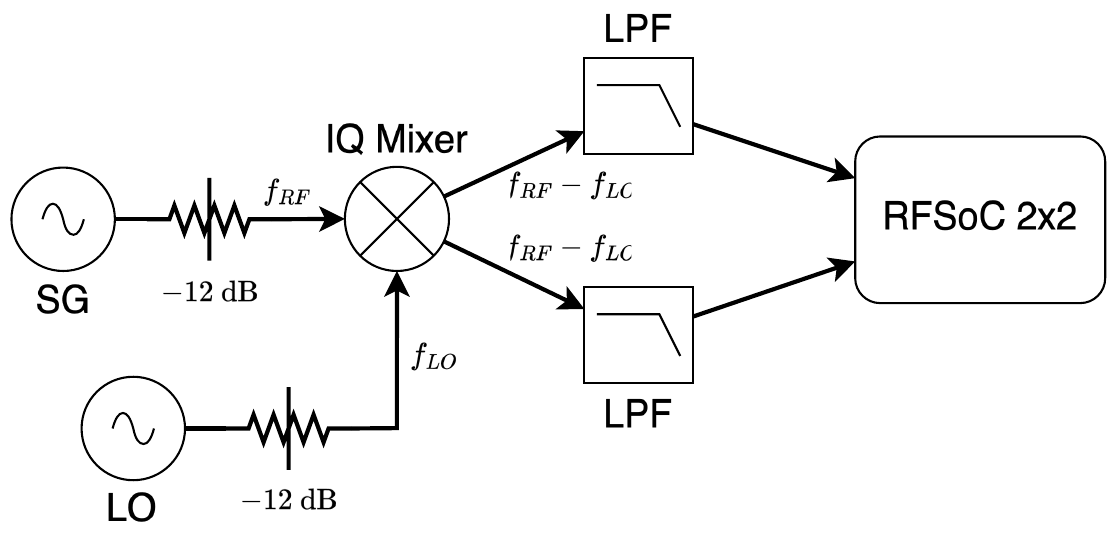}
	\caption{Test setup for spectroscopy and linearity measurement. The input signal (RF) (\SI{16}{GHz}) and LO signal (\SI{15}{GHz}) were generated using signal generators, and attenuated by \SI{12}{dB}.}
	\label{fig_demo_setup}
\end{figure}

We conducted a basic operational check by observing a monotone RF signal with the developed spectrometer.
Input and LO signals were generated using a signal generator (NATIONAL INSTRUMENTS FSL-0020), as shown in \Cref{fig_demo_setup}.
The input signal was set to \SI{16}{GHz}, whereas the LO frequency was set to \SI{15}{GHz}, resulting in an intermediate frequency (IF) signal of \SI{1}{GHz} input to the RFSoC.
\label{sec-test}
\subsection{Demonstration of Spectrometry Operation Using a Monotone RF Source}

\begin{figure}
    \begin{tabular}{cc}
      \begin{minipage}[t]{0.45\hsize}
        \centering\includegraphics[width=2.5in]{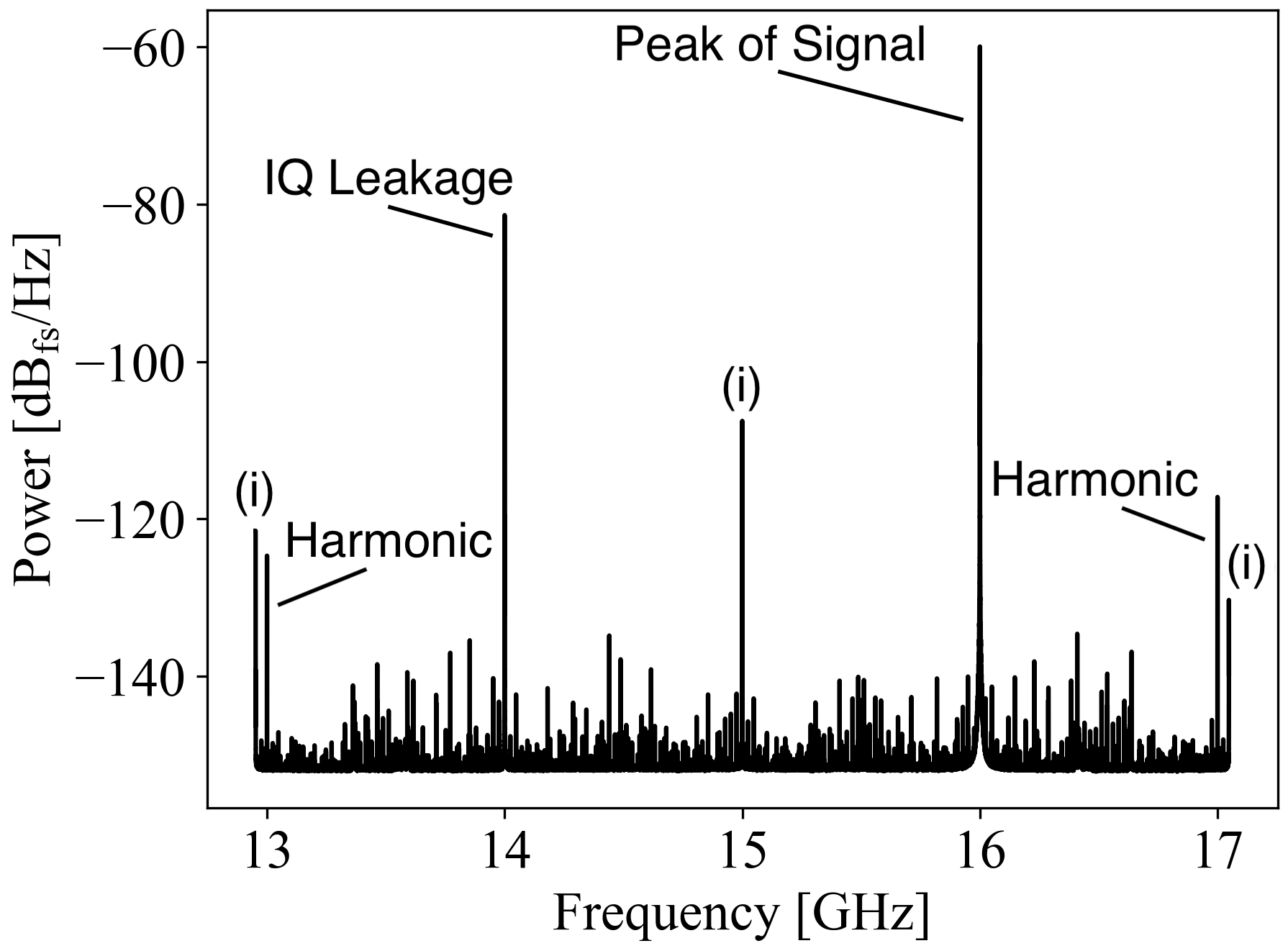}
        \subcaption{}
        \label{fig_demo_wide}
      \end{minipage} &
      \begin{minipage}[t]{0.45\hsize}
        \centering\includegraphics[width=2.5in]{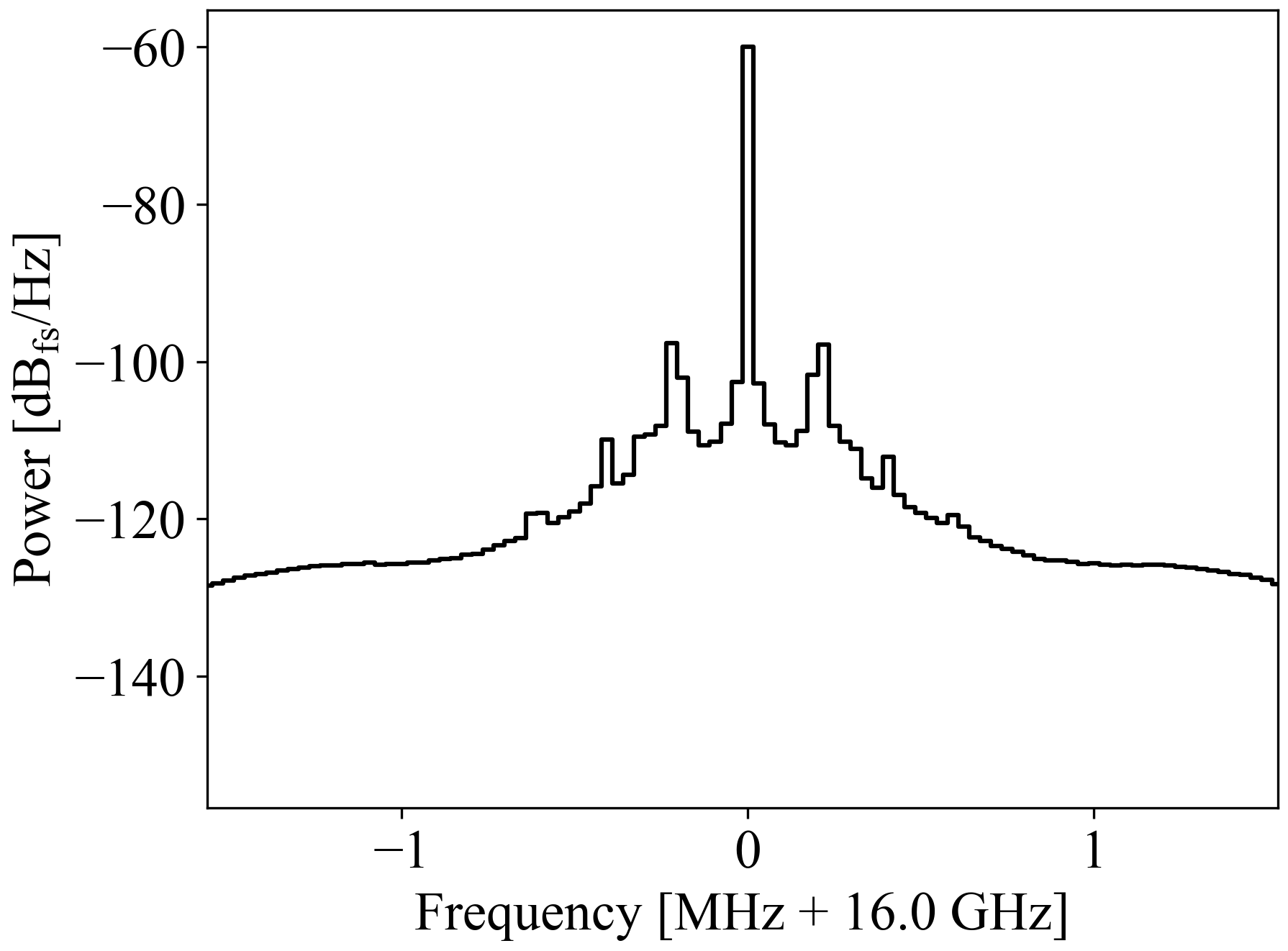}
        \subcaption{}
        \label{fig_demo_narrow}
      \end{minipage} \\
    \end{tabular}
\caption{Power spectrum when a \SI{16}{GHz} signal was input. (a) Complete power spectrum. The peak was accurately observed at \SI{16}{GHz}, with all other peaks being at least \SI{20}{dB} lower. The noise floor was approximately $\sim \SI{-150}{dB_{fs} / Hz}$. (b) Power spectrum near the peak. The peak was accurately located at \SI{16}{GHz}, and the leakage was at least \SI{30}{dB} lower.}
\label{fig_demo}
\end{figure}
The observed spectrum is shown in \Cref{fig_demo_wide}, where the peak was accurately observed in the \SI{16}{GHz} bin.
The leakage from the IQ mixer imbalance and harmonics were all at least \SI{20}{dB} compared with the signal, aligning with the catalog values of the IQ mixer.
The noise floor was approximately $\sim \SI{150}{{dB_{fs} / Hz}}$, also consistent with ADC catalog specifications, thus validating the absence of additional noise in the FFT circuit on the FPGA.

In \Cref{fig_demo_narrow}, an enlarged view of the \SI{16}{GHz} signal reveals symmetric sidelobes around the center, attributable to the DFT calculated within a finite interval and several data points.
This observation aligns with expectations when utilizing a rectangular window (i.e., no window function).

The spectrum exhibits spurious peaks originating from the following sources:
\begin{enumerate}[(i) ]
  \item Components originating from $1/f$ noise near DC (\SI{0}{MHz} and \SI{2048}{MHz})
  \item Leakage of (i) into parallel-computed components (\SI{512}{MHz}, \SI{1024}{MHz}, \SI{1536}{MHz}, ...)
  \item Clock frequency \SI{40.96}{MHz} and its harmonics
\end{enumerate}
The spurious results from (i) are shown in \Cref{fig_demo_wide}.
To mitigate the impact of these spurious peaks, dark matter search experiments should be conducted using multiple LO frequencies.

For the evaluations conducted, a total bandwidth of \SI{50}{MHz} was excluded: \SI{10}{MHz} width for each (i), \SI{1}{MHz} width for each (ii), and \SI{0.25}{MHz} width for each (iii).

\subsection{Linearity}
Subsequently, the linearity of the spectrometer response was assessed. A monotone signal at \SI{14}{GHz} was generated using a signal generator (KEYSIGHT E8257D) at $0.05 \text{--} 2.00 \ \si{mW}$, which was then attenuated by \SI{-60}{dB}.
This power level was adjusted to match that of the previous experiment~\cite{Kotaka2023}, and the LO frequency was set to \SI{13}{GHz}.

\begin{figure}
   \centering\includegraphics[width=4.5in]{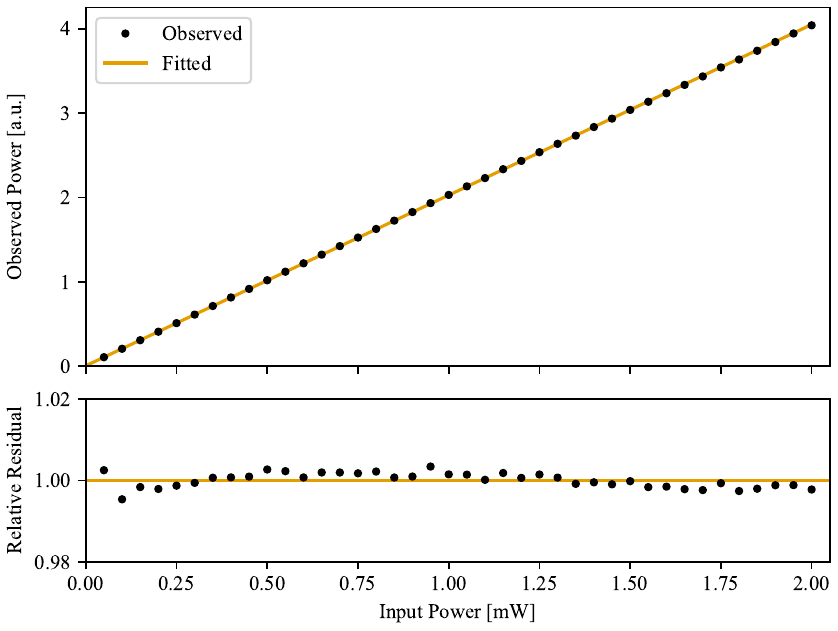}
   \caption{Relationship between input and observed power. The linearity was validated with an accuracy within 0.5\%.}
   \label{fig_linear}
\end{figure}

The relationship between the input and observed power values is shown in \Cref{fig_linear}. 
The obtained power values were fitted to a linear function with an offset and the residuals were calculated from the linear line. We validated the linearity with an accuracy within 0.5\%.
The coupling constant, $\chi$ or $g_{a\gamma}$, is directly proportional to the square root of the observed power. Consequently, the impact of the non-linearity on the coupling constant is less than 0.25\%.

\subsection{Stability and Time Efficiency}
\label{ch-teff}
To assess the stability of the developed system, we conducted a 24-h continuous operation test.
Data were set to be transmitted from the FPGA every \SI{65.536}{ms}, and the system was configured to receive data continuously for 1,318,360 cycles, equivalent to precisely 24 h.
The total time required was 24 h, 44 s, demonstrating that data acquisition was successful more than 99.94\% of the time.

Furthermore, we conducted tests to ensure that each \SI{65.536}{ms} measurement interval was executed with no inefficiencies. This was achieved by analyzing the power fluctuations.
The power $P$ and power fluctuation $\Delta P$ observed when measuring blackbody radiation with a radio-wave receiver connected to the developed spectrometer
in the Rayleigh-Jeans regime can be as follows~\cite{wilson2009tools}:
\begin{align}
   \label{power}
   P &= Gk_B T_{\mathrm{sys}} \Delta \nu, \\
   \label{power_fluc}
   \Delta P &= \frac{Gk_B T_\mathrm{sys} \Delta \nu}{\sqrt{\epsilon \Delta t\Delta \nu }},
\end{align}
where $G$ denotes the system gain, $T_\mathrm{sys}$ represents the system temperature primarily determined from the blackbody temperature and thermal noise from the amplifier, $\Delta \nu$ represents the width of the observed frequency band, $\Delta t$ represents the measurement time, and $\epsilon$ represents the time efficiency of the observation.
This property enables the calculation of $\epsilon$ as follows:
\begin{align}\label{eq:epsilon}
   \epsilon &= \frac{1}{\Delta \nu \Delta t} \left( \frac{P}{\Delta P} \right)^2.
\end{align}

\begin{figure}
   \centering\includegraphics[width=4in]{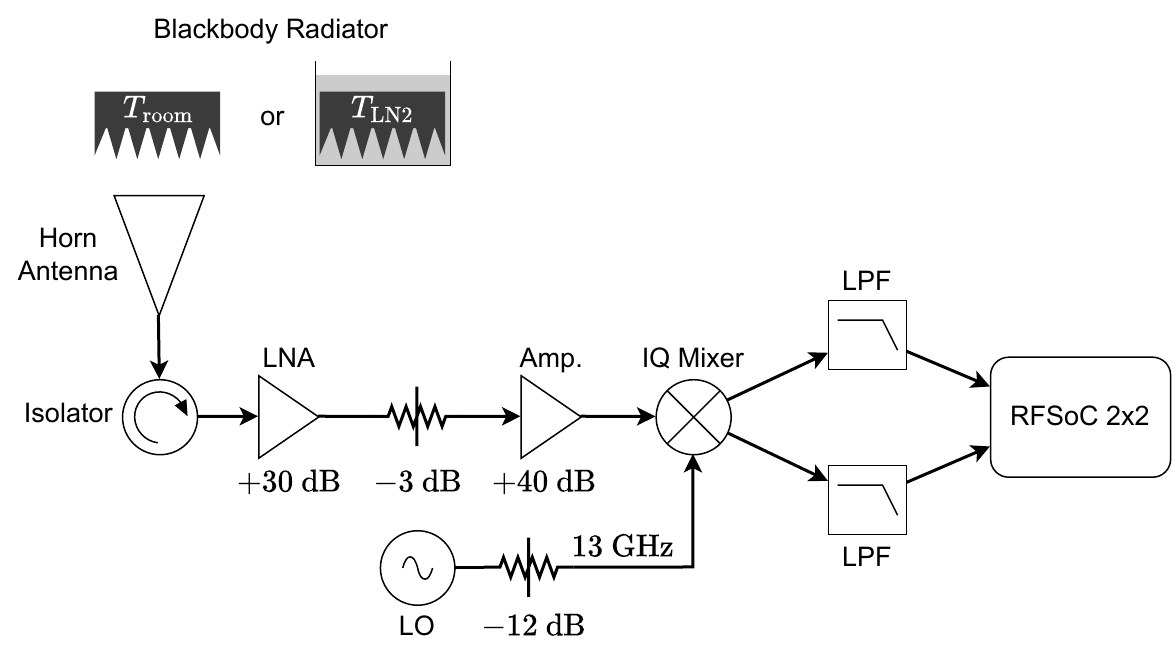}
   \caption{Setup for measuring time efficiency. The radiation from the blackbody was received by the horn antenna. The signal was amplified using two amplifiers, downconverted by an IQ mixer with an LO frequency of \SI{13}{GHz}, and then spectrometered on the RFSoC 2x2 board ranging from $11 \text{--} 15 \ \si{GHz}$. Power spectra were measured for the blackbody at room temperature and the blackbody at \SI{77}{K} immersed in liquid nitrogen.}
   \label{fig_setup_BB}
\end{figure}
The measurement setup is shown in \Cref{fig_setup_BB}.
We received radiation from a blackbody radiator at room temperature (\SI{13}{\degreeCelsius}) or at \SI{77}{K}, achieved by immersing it in liquid nitrogen.
The power spectra were then measured using our spectrometer.
The CV-3 model from E\&C Engineering was utilized for the blackbody radiator, whereas the PE9856/SF-20 model from PASTERNACK was utilized for the horn antenna.
An isolator (PASTERNACK PE8304) was inserted between the horn antenna and the first amplifier (low noise factory LNF-LNR6\_20A), followed by the second amplifier (Mini-Circuits ZVA-18G-S+).
An attenuator was inserted between the first and second amplifiers for gain adjustments.
The spectrometer was configured with an LO frequency of \SI{13}{GHz}, and each measurement lasted for \SI{65.5}{s}, with intervals of \SI{65.5}{ms} repeated 1000 times at both temperatures.
Frequency bands affected by external noise from the folding of the Wi-Fi were excluded from the analysis.

\begin{figure}
    \centering
    \includegraphics[width=2.5in]{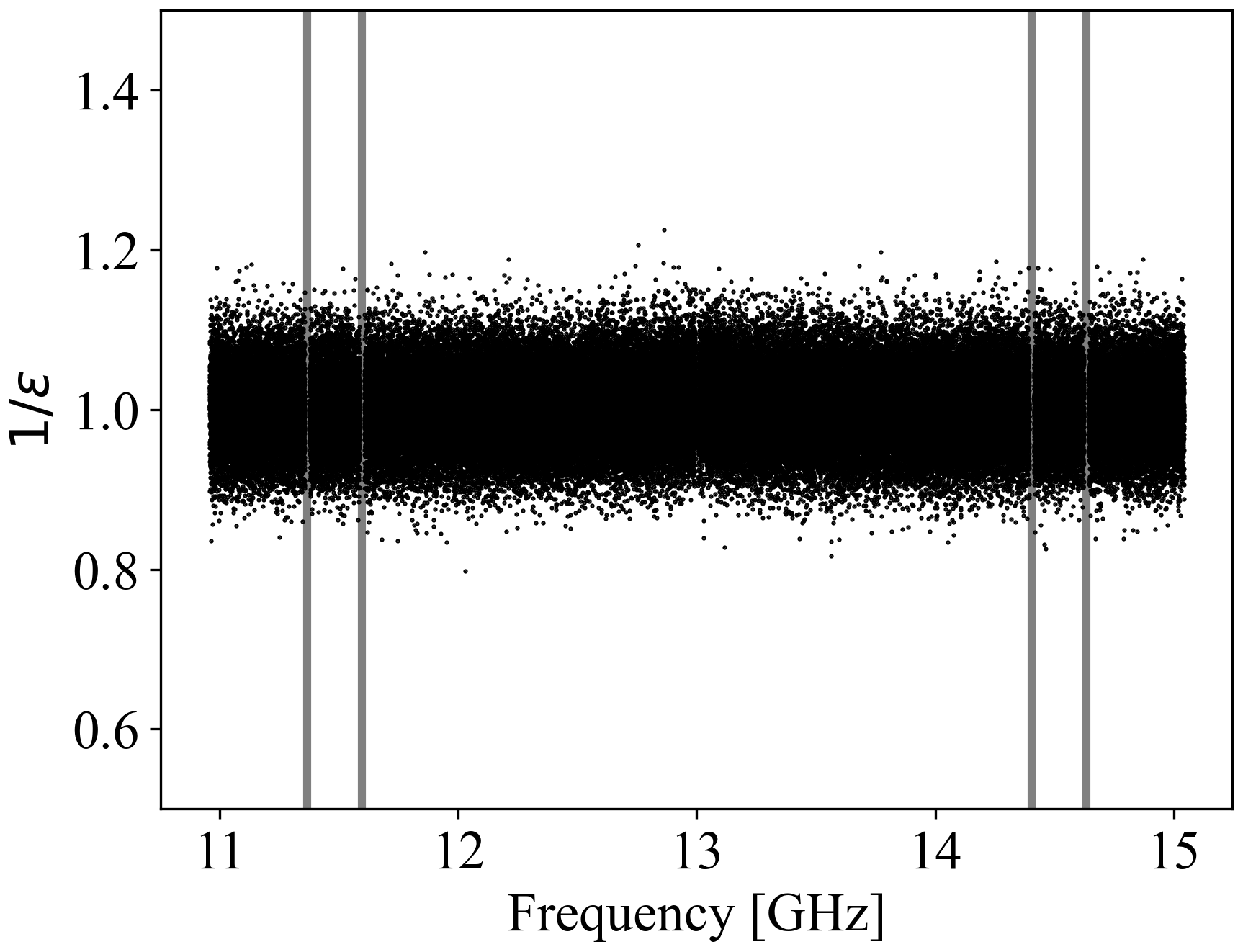}
    \includegraphics[width=2.5in]{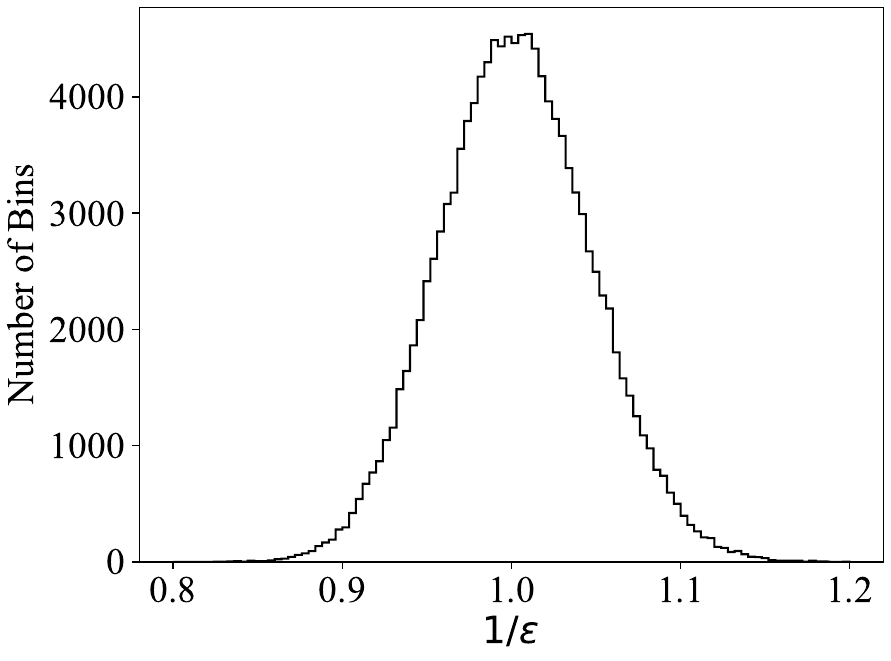}
	\caption{(Left) Measured values of $1 / \epsilon$ at each frequency for room-temperature blackbody radiation. The frequency bands impacted by the folding noise of the Wi-Fi (represented by the thick gray lines) were excluded. (Right) Histogram of the measured $1/\epsilon$.
    The distribution was closely centered around unity, indicating a spectrometer efficiency of approximately 100\%.
    Similar outcomes were observed during the blackbody measurement at \SI{77}{K}.}
	\label{fig_teff}
\end{figure}

When $\Delta P$ was assumed to be Gaussian white noise, $1 / \epsilon$ followed a chi-squared distribution from \Cref{eq:epsilon}.
Instead of directly calculating $\epsilon$, the unbiased $1 / \epsilon$ must be calculated first, followed by the computation of $\epsilon$ from the outcome (\Cref{ap-teff}).
Therefore, for each frequency, we calculated the time average of the power spectrum, fluctuation of the measured values, and subsequently computed $1 / \epsilon (\nu)$. Next, $1 / \epsilon$ was obtained from the average of these values.
The values of $1 / \epsilon$ obtained for each frequency and its histogram are shown in \Cref{fig_teff}.
From the measurement of room-temperature blackbody radiation, we obtained $1 / \epsilon = \num{1.00257(13)}$, and from the measurement of the \SI{77}{K} blackbody, we obtained $1 / \epsilon = \num{1.00169(13)}$.
Both measurements ensured that $\epsilon$ exceeded 99.7\%, thus validating the spectrometer's capability to perform spectrometry with better than 99.6\% time efficiency.

\section{Conclusion}
\label{sec-conclusion}
This study developed a wide-band spectrometer that enabled spectrometry across a \SI{4}{GHz} bandwidth, focusing on the need to search for axions and dark photons.
Additionally, a frequency resolution of \SI{16}{kHz} was achieved to detect the extremely narrow spectral peak of the conversion photon, typically $\Delta \nu / \nu \sim 10^{-6}$.
With this spectrometer, we anticipate conducting an efficient search and expect an enhancement in the sensitivity of $\chi$ or $g_{a\gamma\gamma}$ owing to increased data statistics.

To achieve this performances, we designed the hardware and firmware components, including the logic circuits on the FPGA.
To digitize IQ signals, we utilized two ADCs and an IQ mixer, and by incorporating two \SI{4}{GSPS} ADCs within the RFSoC, we secured a bandwidth of \SI{4}{GHz}.
Furthermore, on the built-in FPGA, we implemented FFT conversions through 16-parallel FFT circuits and utilized an accumulator circuit to compress the data, allowing for signal analysis from the ADC with no loss.
To minimize the increase in resource consumption of FPGA, resulting from parallelization, we developed a memory-efficient rotation method and implemented a Radix-$2^4$ architecture that required fewer general rotation units.
These enhancements resulted in the development of a spectrometer with the required performance.

Performance evaluation tests were conducted on the developed spectrometer, which included obtaining the desired spectrum by inputting a monotone RF source and validating linearity at an accuracy within 0.5\%.
Additionally, we validated the stability of the spectrometer for continuous operation over 24 h and demonstrated its ability to perform spectrometry with greater than 99.6\% time efficiency.

The increased statistics provided by this spectrometer will significantly enhance the sensitivity of axion and dark photon searches.
Although previous experiments~\cite{Kotaka2023, Adachi2024} were limited to searching a \SI{2}{MHz} frequency range at once, our spectrometer allows for a \SI{4}{GHz} search range, resulting in a time efficiency improvement of over $2000$ times.
Additionally, our approach achieves approximately twice the search efficiency compared to conventional approaches~\cite{Bajjali2023, Klein2012}.
Moreover, the developed spectrometer has potential in various fields, including radio astronomy and industry.
Furthermore, the current FPGA implementation utilizes fewer than half of the available resources, which allows for potential bandwidth doubling by implementing two FFT circuits and utilizing four ADCs in the next-generation boards Ref. \cite{RFSoC4x2}.

\section*{Acknowledgment}

This research was supported by JSPS KAKENHI under grant numbers 22K18713, 23H00111, 24KK0066, and 24KJ1452,
as well as support from JST, CREST under grant number JPMJCR23I3.
Additionally, this research was supported by grants from the Murata and Sumitomo Foundations.
SA acknowledges the Hakubi Project Program of Kyoto University.

\appendix

\section{Estimation Method for Time Efficiency $\epsilon$}
\label{ap-teff}
\Cref{ch-teff} delves into time efficiency $\epsilon$, which does not adhere to a normal distribution.
Therefore, the precise estimation of $\epsilon$ requires care.
Hence, this section explains the method of its estimation.

As shown in \Cref{ch-teff}, the measured value $P_\mathrm{obs} (\nu, t)$, of the thermal noise, $P$, measured within the frequency range $\nu$ to $\nu + \Delta \nu$ and time range $t$ to $t + \Delta t$, conforms to a normal distribution with a mean and standard deviation of $P$ and $\Delta P$, respectively.
Additionally, $P$ and $\Delta P$ are theoretically expected to follow \Cref{power,power_fluc}.

When $P_\mathrm{obs} (\nu, t)$ is measured $n_t$ times with respect to $t$, $P(\nu)$ can be estimated through the (time) average $\overline{P_\mathrm{obs}} (\nu)$ of $P_\mathrm{obs} (\nu, t)$, whereas $\Delta P(\nu)^2$ can be estimated using the unbiased variance, $V(P_\mathrm{obs})(\nu)$, of $P_\mathrm{obs} (\nu, t)$ with respect to $t$.
Given that $P_\mathrm{obs} (\nu, t)$ adheres to a normal distribution with variance $P(\nu)^2 / (\Delta \nu \epsilon \Delta t)$, $V(P_\mathrm{obs})(\nu)^2 \cdot \Delta \nu \epsilon \Delta t / P(\nu)^2$ conforms to a chi-squared distribution with $n_t - 1$ degrees of freedom.
Therefore, defining a random variable, $X_\mathrm{obs}(\nu)$, as $X_\mathrm{obs}(\nu) = V(P_\mathrm{obs})(\nu)^2 \cdot \Delta \nu \Delta t / (P(\nu)^2 \cdot (n_t - 1))$, the expectation $E$ and the unbiased variance, $V$, of $X_\mathrm{obs}(\nu)$ can be expressed as follows:
\begin{align}
   E(X_\mathrm{obs}) &= \frac{1}{\epsilon} \\
   V(X_\mathrm{obs}) &= \frac{2}{(n_t-1)\epsilon^2}
\end{align}
Therefore, $1 / \epsilon$ can be estimated from the measurement of $X_\mathrm{obs}$.


%

\vspace{0.2cm}
\noindent


\let\doi\relax


\bibliographystyle{ptephy}
\bibliography{main}

\end{document}